\documentclass[%
prl,
 reprint,
superscriptaddress,
nofootinbib,
 amsmath,amssymb,
 aps,
]{revtex4-2}

\usepackage{verbatim}
\usepackage{scrextend}
\usepackage{listings}
\usepackage{graphicx}
\usepackage{dcolumn}
\usepackage{lineno} 
\usepackage{bm}
\usepackage{subfigure}
\usepackage{units}

\usepackage{hyperref}
\hypersetup{
    colorlinks,
    linkcolor={blue!50!black},
    citecolor={blue!50!black},
    urlcolor={blue!50!black},
}

\usepackage[dvipsnames]{xcolor}

\begin{document}

\title{Search for Light Dark Matter with NEWS-G at the Laboratoire Souterrain de Modane Using a Methane Target}

\author{M.~M.~Arora}
\affiliation{Department of Mechanical and Materials Engineering, Queen’s University, Kingston, Ontario K7L 3N6, Canada}
\author{L.~Balogh}
\affiliation{Department of Mechanical and Materials Engineering, Queen’s University, Kingston, Ontario K7L 3N6, Canada}
\author{C.~Beaufort}
\affiliation{LPSC-LSM, Universit\'{e} Grenoble-Alpes, CNRS-IN2P3, Grenoble 38026, France}
\author{A.~Brossard}
\altaffiliation[now at ]{TRIUMF, Vancouver, BC V6T 2A3, Canada}
\affiliation{Department of Physics, Engineering Physics, and Astronomy, Queen’s University, Kingston, Ontario K7L 3N6, Canada}
\author{M.~Chapellier}
\affiliation{Department of Physics, Engineering Physics, and Astronomy, Queen’s University, Kingston, Ontario K7L 3N6, Canada}
\author{J.~Clarke}
\affiliation{Department of Physics, Engineering Physics, and Astronomy, Queen’s University, Kingston, Ontario K7L 3N6, Canada}
\author{E.~C.~Corcoran}
\affiliation{Chemistry and Chemical Engineering Department, Royal Military College of Canada, Kingston, Ontario K7K 7B4, Canada}
\author{J.-M.~Coquillat}
 \email[e-mail:]{jeanmarie.coquillat@queensu.ca}
\affiliation{Department of Physics, Engineering Physics, and Astronomy, Queen’s University, Kingston, Ontario K7L 3N6, Canada}
\author{A.~Dastgheibi-Fard}
\affiliation{LPSC-LSM, Universit\'{e} Grenoble-Alpes, CNRS-IN2P3, Grenoble 38026, France}
\author{Y.~Deng}
\affiliation{Department of Physics, University of Alberta, Edmonton, Alberta T6G 2E1, Canada}
\author{D.~Durnford}
 \email[e-mail:]{ddurnfor@ualberta.ca}
\affiliation{Department of Physics, University of Alberta, Edmonton, Alberta T6G 2E1, Canada}
\author{C.~Garrah}
\affiliation{Department of Physics, University of Alberta, Edmonton, Alberta T6G 2E1, Canada}
\author{G.~Gerbier}
\affiliation{Department of Physics, Engineering Physics, and Astronomy, Queen’s University, Kingston, Ontario K7L 3N6, Canada}
\author{I.~Giomataris}
\affiliation{IRFU, CEA, Universit\'{e} Paris-Saclay, F-91191 Gif-sur-Yvette, France}
\author{G.~Giroux}
\affiliation{Department of Physics, Engineering Physics, and Astronomy, Queen’s University, Kingston, Ontario K7L 3N6, Canada}
\author{P.~Gorel}
\affiliation{SNOLAB, Lively, Ontario P3Y 1N2, Canada}
\author{M.~Gros}
\affiliation{IRFU, CEA, Universit\'{e} Paris-Saclay, F-91191 Gif-sur-Yvette, France}
\author{P.~Gros}
\affiliation{Department of Physics, Engineering Physics, and Astronomy, Queen’s University, Kingston, Ontario K7L 3N6, Canada}
\author{O.~Guillaudin}
\affiliation{LPSC, Universit\'{e} Grenoble-Alpes, CNRS-IN2P3, Grenoble 38026, France}
\author{E.~W. Hoppe}
\affiliation{Pacific Northwest National Laboratory, Richland, Washington 99354, USA}
\author{I.~Katsioulas}
\altaffiliation[now at ]{European Spallation Source ESS ERIC (ESS), Lund, SE-221 00, Sweden}
\affiliation{School of Physics and Astronomy, University of Birmingham, Birmingham B15 2TT, United Kingdom}
\author{F.~Kelly}
\affiliation{Chemistry and Chemical Engineering Department, Royal Military College of Canada, Kingston, Ontario K7K 7B4, Canada}
\author{P.~Knights}
\affiliation{School of Physics and Astronomy, University of Birmingham, Birmingham B15 2TT, United Kingdom}
\author{P.~Lautridou}
\affiliation{SUBATECH, IMT-Atlantique/CNRS-IN2P3/Nantes University, Nantes, 44307, France}
\author{A.~Makowski}
\affiliation{Department of Physics, Engineering Physics, and Astronomy, Queen’s University, Kingston, Ontario K7L 3N6, Canada}
\author{I.~Manthos}
\affiliation{School of Physics and Astronomy, University of Birmingham, Birmingham B15 2TT, United Kingdom}
\affiliation{Institute for Experimental Physics, University of Hamburg, Hamburg 22767, Germany}
\author{R.~D.~Martin}
\affiliation{Department of Physics, Engineering Physics, and Astronomy, Queen’s University, Kingston, Ontario K7L 3N6, Canada}
\author{J.~Matthews}
\affiliation{School of Physics and Astronomy, University of Birmingham, Birmingham B15 2TT, United Kingdom}
\author{H.~M.~McCallum}
\affiliation{Chemistry and Chemical Engineering Department, Royal Military College of Canada, Kingston, Ontario K7K 7B4, Canada}
\author{H.~Meadows}
\affiliation{Department of Physics, Engineering Physics, and Astronomy, Queen’s University, Kingston, Ontario K7L 3N6, Canada}
\author{L.~Millins}
\affiliation{School of Physics and Astronomy, University of Birmingham, Birmingham B15 2TT, United Kingdom}
\altaffiliation[also at]{Particle Physics Department, STFC Rutherford Appleton Laboratory, Chilton, Didcot, OX11 OQX, UK}
\author{J.-F.~Muraz}
\affiliation{LPSC-LSM, Universit\'{e} Grenoble-Alpes, CNRS-IN2P3, Grenoble 38026, France}
\author{T.~Neep}
\affiliation{School of Physics and Astronomy, University of Birmingham, Birmingham B15 2TT, United Kingdom}
\author{K.~Nikolopoulos}
\affiliation{School of Physics and Astronomy, University of Birmingham, Birmingham B15 2TT, United Kingdom}
\affiliation{Institute for Experimental Physics, University of Hamburg, Hamburg 22767, Germany}
\author{N.~Panchal}
\affiliation{Department of Physics, Engineering Physics, and Astronomy, Queen’s University, Kingston, Ontario K7L 3N6, Canada}
\author{M.-C.~Piro}
\affiliation{Department of Physics, University of Alberta, Edmonton, Alberta T6G 2E1, Canada}
\author{N.~Rowe}
\affiliation{Department of Physics, Engineering Physics, and Astronomy, Queen’s University, Kingston, Ontario K7L 3N6, Canada}
\author{D.~Santos}
\affiliation{LPSC-LSM, Universit\'{e} Grenoble-Alpes, CNRS-IN2P3, Grenoble 38026, France}
\author{G.~Savvidis}
\affiliation{Department of Physics, Engineering Physics, and Astronomy, Queen’s University, Kingston, Ontario K7L 3N6, Canada}
\author{I.~Savvidis}
\affiliation{Aristotle University of Thessaloniki, Thessaloniki 54124, Greece}
\author{D.~Spathara}
\affiliation{School of Physics and Astronomy, University of Birmingham, Birmingham B15 2TT, United Kingdom}
\author{F.~Vazquez~de~Sola~Fernandez}
 \email[e-mail:]{14favd@queensu.ca}
 \altaffiliation[now at ]{Nikhef (Nationaal instituut voor subatomaire fysica), Science Park 105, 1098 XG Amsterdam, Netherlands}
\affiliation{SUBATECH, IMT-Atlantique/CNRS-IN2P3/Nantes University, Nantes, 44307, France}
\author{R.~Ward}
\affiliation{School of Physics and Astronomy, University of Birmingham, Birmingham B15 2TT, United Kingdom}
\affiliation{Institute for Experimental Physics, University of Hamburg, Hamburg 22767, Germany}

\collaboration{NEWS-G Collaboration}

\date{\today}

\begin{abstract}
The NEWS-G direct detection experiment uses spherical proportional counters to search for light dark matter candidates. New results from a 10 day physics run with a $135~\mathrm{cm}$ in diameter spherical proportional counter at the Laboratoire Souterrain de Modane are reported. The target consists of $114~\mathrm{g}$ of methane, providing sensitivity to dark matter spin-dependent coupling to protons. New constraints are presented in the mass range $0.17$ to $1.2~\mathrm{GeV/c^2}$, with a 90\% confidence level cross-section upper limit of $30.9~\mathrm{pb}$ for a mass of $0.76~\mathrm{GeV/c^2}$. 
\end{abstract}

\maketitle

Astronomical and cosmological observations strongly suggest the existence of nonbaryonic dark matter (DM) in our Universe~\cite{Clowe_2006,plank}. Theories beyond the standard model provide DM candidates in the form of nonrelativistic, weakly interacting massive particles (WIMPs) \cite{snowmass2021,PDG, feng} constituting our galactic halo. Beyond the canonical prediction of thermal relic WIMPs with masses ranging from approximately $10$ to $1000$ $\mathrm{GeV/c^2}$, recent theoretical models such as asymmetric DM \cite{asymmetric1,asymmetric2, Balan:2024cmq} favor lighter DM, possibly at or below $1~\mathrm{GeV/c^2}$ \cite{Essig2013,sector}. Such $\mathcal{O}(1)~\mathrm{GeV/c^2}$ DM candidates may still be produced by the thermal freeze-out mechanism proposed for traditional WIMP DM if new force mediators exist \cite{bozorgnia2024darkmattercandidatessearches}, motivating experimental searches \cite{snowmass_lowmass, appec}.

The NEWS-G experiment employs spherical proportional counters (SPCs) filled with various gases to search for WIMP-like particles scattering off target nuclei, producing nuclear recoil energies of up to several keV. First constraints using this technology were obtained with a $60~\mathrm{cm}$ in diameter SPC filled with a neon-methane mixture~\cite{newsg_first}. The new S140 detector~\cite{Brossard2022} consists of a grounded 135~cm in diameter SPC constructed with low background oxygen-free high-conductivity copper (C10100) featuring a 0.5~mm inner shield made of high-purity electroformed copper~\cite{NEWS-G:2020fhm}. It is equipped~\cite{Brossard2022} with the multianode sensor ``ACHINOS''~\cite{Giganon:2017isb, Giomataris_2020}, which features 11 anodes. The sensor is developed to optimize both ionization electron collection and high charge amplification capabilities. It is held at the center of the SPC with a support rod, which also carries the wires used to apply a high voltage on the anodes and from which the signal is read out.

This Letter reports new results using methane. This hydrogen-rich target benefits from a favorable kinematic matching to WIMP-like DM candidates with $\mathcal{O}\left(1~\mathrm{GeV/c^2} \right)$ masses, increasing momentum transfer and hence the detector's sensitivity. Furthermore, the unpaired proton provides sensitivity to spin-dependent (SD) DM-proton couplings. The data were collected during the commissioning of the S140 at a depth of 4800 meters of water equivalent in the Laboratoire Souterrain de Modane \cite{lsm_depth} in October 2019, prior to its transportation to SNOLAB. The SPC was operated at $135~\mathrm{mbar}$ of pressure (equivalent to $114~\mathrm{g}$) surrounded with a water shield for 10.26 days. After accounting for dead time and data set aside for training---which have previously been used to produce a preliminary DM exclusion limit \cite{IDM_newsg}---an exposure of $0.599~\mathrm{kg~d}$ was used to obtain the final physics results presented here.


After an energy deposition in the gas volume, ionization electrons drift toward the anodes, where they are multiplied by an avalanche process close to the anode. The measured signal from a single ionization electron reaching the anode is a combination of the current induced by the avalanche ions as they drift toward the cathode, and the response of a charge-sensitive preamplifier. 

For events above approximately 30 ionization electrons, which include some of the calibrations used in this work, the event amplitude, defined as the integral of the pulse, is used as an estimator of the detected event's energy. Additionally, for pointlike energy depositions, the 10\% to 90\% rise time of the integrated pulse is proportional to the diffusion experienced by ionization electrons. As this increases with the radial position of the event, selection cuts on the rise time are used to reject surface background events. For events with fewer than approximately 10 ionization electrons, the pulses are deconvolved to remove both the preamplifier response and the time structure of the current induced by the avalanche ions. Thus, a series of delta impulses is obtained, each corresponding to the arrival of an ionization electron to the anodes~\cite{Vazquez2020Thesis_PFsection}. Each of these impulses will have a different amplitude, proportional to the size of the corresponding avalanche.

The relatively large longitudinal diffusion of the electrons in methane results in $\mathcal{O}(100~\mathrm{\mu s})$ spread in their arrival times. This allows for the identification of individual ionization electrons in processed traces for low energy events, such as those generated by low-mass WIMP recoils. A peak-finding (PF) algorithm based on the ROOT TSpectrum search method \cite{root, ROOT_TSpectrum} is applied to estimate the number of electrons present in the waveform, their arrival times, and avalanche size. The number of observed peaks, which is linked to deposited energy, constitutes the first parameter on which the present analysis is based. Although this is connected to the number of ionization electrons generated in the event, a fraction of them may be lost in baseline noise fluctuations, or electrons arriving in close temporal proximity might not be resolvable. The second parameter is the time separation between the first and the last peak. This parameter is used to statistically discriminate contributions from different background sources.

The overall principle of the analysis is based on the comparison of the time separation distributions for two, three, and four observed peaks in data, and the expected backgrounds and signal. The 11 ACHINOS anodes are grouped into two readout channels: one comprising the five nearest anodes to the support rod (``near'' anodes) and the other with the six farthest (``far'' anodes). Only events collected from the far hemisphere, where the electric field is more homogeneous, are considered, defining the analysis fiducial volume.
The relative signal of the two readout channels is used as an event quality selection. Electron multiplication occurring near an anode induces also a smaller signal of opposite polarity to the other anodes, as expected by the Shockley-Ramo theorem~\cite{Shockley,Ramo} and discussed in Ref.~\cite{Katsioulas:2022cqe}. The absence of this cross-channel signal for an event localized at one anode suggests it did not originate from ionization electron amplification in the detector, and hence was removed (``antispikes cut''). Following an $\alpha$ decay on the surface of the detector, there is a transient increase in the rate of single-electron events. Therefore, events occurring within $5$~s after every $\alpha$-particle detection were also discarded. The collected dataset is separated into two subsets: 27\% is the ``training data'' on which the analysis is tuned, and the remaining 73\% is ``DM search data'' on which the search is performed. 

Extensive calibrations of the detector response were carried out for this physics campaign. A UV laser calibration system was implemented as described in Ref.~\cite{newsg_laser}. The laser output is separated in two parts with an optical fiber splitter. One part is fed into the active volume of the SPC with an optical fiber. The bare end of the optical fiber is directed at the far hemisphere of the SPC from the inner surface of which photoelectrons are extracted. The other part is directed to a photodiode, whose signal is used to identify laser-induced events in the detector.

During data collection, the UV laser was operated at a high intensity to monitor changes in the detector response in real time. This revealed that over the course of physics data collection, the detector gas gain was reduced by approximately $11\%$ due to deterioration of the gas quality during operation in sealed mode (see Appendix B). When operated at low intensity, the UV laser can induce a signal dominated by single photoelectron events, which is used to model the detector response. The electron avalanche multiplication is modeled with the Polya distribution~\cite{L11,L12,L13,L14,newsg_laser} with shape parameter $\theta$, scaled to the mean gain of the detector $\left \langle G \right \rangle$ (see Appendix A). During the physics campaign, approximately 1 h of calibration with the UV laser was obtained for each day of data taking, yielding an overall result of $\theta = 0.125^{+0.026}_{-0.023}$, indicating an approximately exponential distribution. Additionally, the UV laser was used to determine the detector trigger efficiency. The trigger algorithm used online to collect physics data was also applied offline to the laser calibration data. It was found that $64_{-3}^{+4}\%$ of single-electron and as much as $93_{-1}^{+2}\%$ of double-electron events fulfill the trigger requirement. For multiple-electron events close to the anodes, the increased electron pileup probability increases the trigger efficiency. 

The performance of the PF method was also evaluated using laser data. On average, 
$62\pm3\%$ 
of single electrons fulfill the amplitude threshold for peak detection. The quoted uncertainty includes the effect of the observed time dependence of the gas gain during the campaign. The minimum time separation required for the PF method to distinguish two electrons was calibrated with laser data events where exactly two peaks were found by the method. The observed rate drops off for events with very short time separation, which was fit with an error function with a threshold of $8.2\pm0.4~\mathrm{\mu s}$. The PF parameters were chosen so as to minimize the probability of baseline noise generating a false peak, to avoid single-electron events being reconstructed as containing two peaks. A probability of 0.03\% per $1.05~\mathrm{ms}$ search window was obtained, allowing this effect to be neglected. From these results, a model was developed to compute the probability that surface events with $n$ ionization electrons and expected diffusion time $\sigma$ will be reconstructed as having $k$ peaks (see Appendix C). The resulting predictions for three- and four-peak events were consistent with the respective laser calibration data.

\begin{table*}[htp!]
\begin{center}
\begin{tabular}{r|cccccccccccccc}
    \hline \hline
    & \multicolumn{14}{c}{Ionization Electrons} \\ \cline{2-15}
   Peaks  & 2 & 3 & 4 & 5 & 6 & 7 & 8 & 9 & 10 & 11 & 12 & 13 & 14 & 15 \\ \hline
    2    & 0.124 & 0.227 & 0.265 & 0.255 & 0.221 & 0.181 & 0.144 & 0.112 & 0.088 & 0.069 & 0.056 & 0.046 & 0.038 & 0.033  \\
    3    & $\cdots$     & 0.049 & 0.117 & 0.174 & 0.206 & 0.213 & 0.203 & 0.183 & 0.158 & 0.133 & 0.111 & 0.092 & 0.077 & 0.064  \\
    4    & $\cdots$     & $\cdots$     & 0.016 & 0.051 & 0.093 & 0.132 & 0.161 & 0.176 & 0.179 & 0.172 & 0.160 & 0.144 & 0.127 & 0.111 \\
    Total     & 0.124 & 0.276 & 0.400 & 0.480 & 0.522 & 0.527 & 0.509 & 0.472 & 0.426 & 0.375 & 0.327 & 0.283 & 0.243 & 0.210 \\ \hline \hline
    \end{tabular}
    \protect\caption{Combined detection and selection efficiency for volume events, depending on the number of ionization electrons.}
    \label{tab:PFeff}
\end{center}
\end{table*}


At the end of the physics campaign, $\mathrm{^{37}Ar}$ was injected into the detector. Decaying via electron capture, $\mathrm{^{37}Ar}$ produces low-energy electrons and x-rays uniformly throughout the detector volume. The dominant total energy depositions per event are $270~\mathrm{eV}$ and $2.8~\mathrm{keV}$ \cite{Ar37_5,newsg_ar37} with smaller contributions from different decay paths and partially escaped decays \cite{darkside_calib}. This calibration data is used to measure \textit{in situ} the overall energy response, including the mean avalanche gain of all six far-channel anodes. The ionization yield $W(E)$ of the detector for interactions induced by electrons and photons, collectively referred to as ``electronic recoils,'' is parametrized using the expression of Ref.~\cite{inokuti}, with free parameters $U$ for the asymptote, and $W_0$ as the corresponding high energy limiting value. Independent measurements of $W_0$ and $U$ were performed~\cite{Arora:2024tbs} and used as a prior for the present calibration, leading to a result of $W_0 = 30.0_{-0.15}^{+0.14}~\mathrm{eV}$ and $U = 15.70^{+0.52}_{-0.34}~\mathrm{eV}$.
This value of $U$ establishes a minimum energy threshold for ionization events to occur, given the adopted energy response model \cite{inokuti}.
The statistical dispersion of ionization is controlled by the Fano factor~\cite{uno2}, whose value was obtained from the literature~\cite{grosswendt_1985} as an exact number, given there were no uncertainties provided. This dispersion was modeled with a COM-Poisson distribution~\cite{mainCOM,newsg_fano} (see Appendix B). Ionization electron losses through attachment are parametrized using this calibration data by assuming a survival probability varying linearly with radial position.

The $^{37}$Ar data was used to characterize electron diffusion as a function of radius, following an empirical relationship of $\sigma(r) = \sigma_{\text{max}}\,(r/r_{\text{max}})^\alpha$, with $\sigma$ being the standard deviation of the electron arrival time, $r$ the radius of the interaction, the subscript $\text{max}$ indicating the value at the cathode surface ($\sigma_{\text{max}}=123.5\pm1.1~\mathrm{\mu s}$ for physics runs), and $\alpha = 3.05\pm0.15$. This diffusion model was combined with the PF performance model to generate Monte Carlo (MC) simulations of two-, three-, and four-peak volume events (see Appendix C).

\begin{figure}[t]
\centering
  \includegraphics[width=0.95\linewidth]{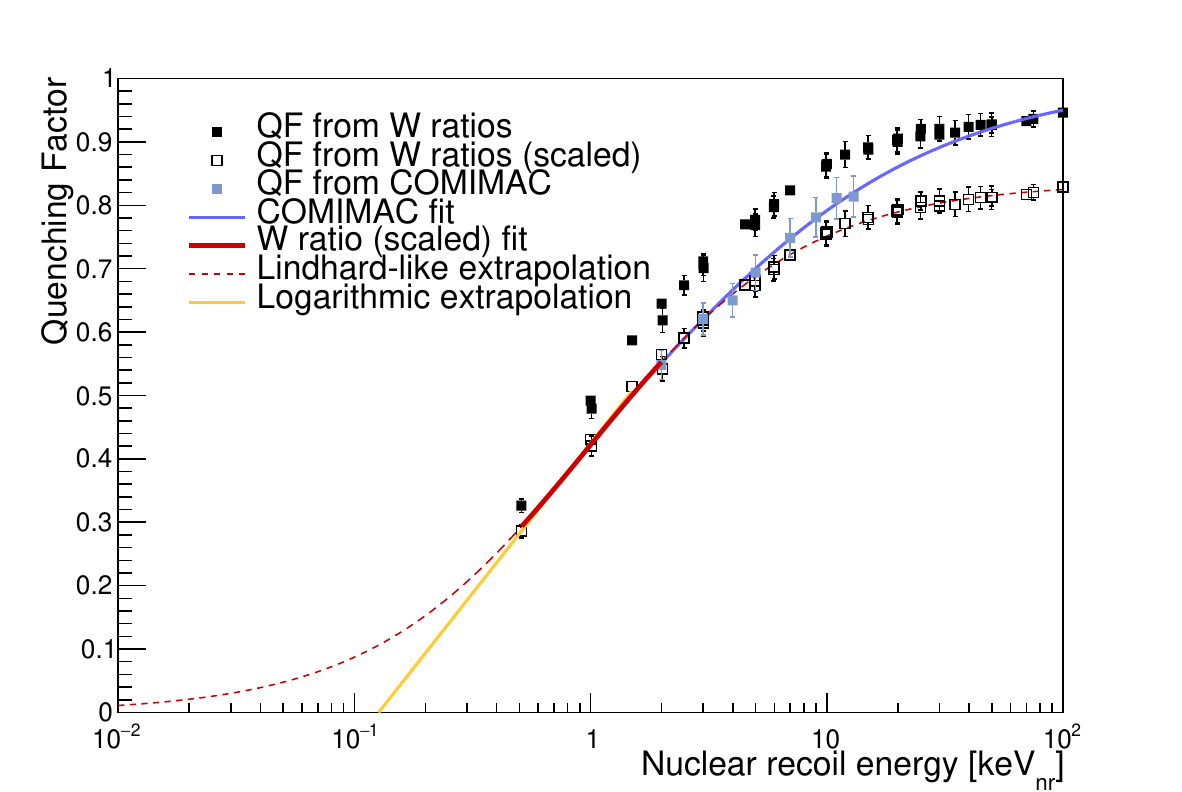}
  \caption{Quenching factor used in this work. The curves used, from highest to lowest energies, are the COMIMAC measurement (lavender line \cite{comimac}, above $2~\mathrm{keV}$), scaled $W$-value ratios (red line \cite{katsioulasQF}, between $2~\mathrm{keV}$ and $510~\mathrm{eV}$), and a logarithmic extrapolation (orange line, below $510~\mathrm{eV}$); the Lindhard-like extrapolation is shown for comparison (dotted red line \cite{osti_4701226}).}
  \label{fig:QF}
\end{figure}

The detector response to nuclear recoils is affected by the ionization quenching factor (QF), defined as the ratio of ionization energies released by a nuclear and an electronic recoil of the same kinetic energy. Two independent measures of the QF for hydrogen atoms in CH$_{4}$ were used in this analysis: (i) a NEWS-G measurement with the COMIMAC facility was performed between $2$ and $13~\mathrm{keV}$ \cite{comimac}, and (ii) a QF curve was estimated from $W$-value measurements in the energy range $510~\mathrm{eV_{nr}}$ to several hundreds of $\mathrm{keV}$~\cite{katsioulasQF}. The latter was conservatively scaled down by $15\%$ to be in agreement with the COMIMAC measurement. Below $510~\mathrm{eV_{nr}}$, the extrapolation $QF(E_{K}) = 0.428 + 0.224\ln(E_{K})$ was used based on the scaled $W$-value curve, effectively setting an energy threshold at $QF(127~\mathrm{eV_{nr}})=0$. This approach is conservative compared to an extrapolation based on the widely used Lindhard model~\cite{osti_4701226} (see Fig.~\ref{fig:QF}). 

The fiducial volume acceptance of the detector’s far channel for $2$ through $15$ electron events was determined as a function of their radial position using an MC simulation of the electron drift \cite{carter_thesis_fid_section}. The finite element method software \textsc{comsol} was used to model the electric field~\cite{comsol}, and the electron transport properties in the gas were obtained using the \textsc{magboltz} computer program~\cite{magboltz}. For example, 2-electron (15-electron) events originating from the cathode surface of the SPC have a $73.83\pm0.04\%$ ($64.72\pm0.05\%$) probability of fulfilling the fiducialization cut. The fiducialization efficiency was validated against $\mathrm{^{37}Ar}$ calibration data \cite{george_thesis_fid_section}. The efficiencies of the additional event selection requirements, were also estimated. The $\alpha$ particle cut results in a 14$\%$ dead time. The antispikes cut removes 95\% of events not associated with an electron avalanche while keeping 77\% of ionization electrons. Removing tagged laser calibration events reduces the effective runtime by $0.48\%$.

After applying the previous selection criteria, a rate of $403~\mathrm{mHz}$ of single-peak events was observed in the training data, compared to only $6.5$ and $1.1~\mathrm{mHz}$ of two- and three-peak events, respectively. In the absence of an explanation for such a high single-peak event rate, it was decided to keep only events with two or more peaks to calculate WIMP exclusion limits. The combined detection and selection cut efficiency for volume events with up to 15 electrons for the physics conditions of this work are shown in Table~\ref{tab:PFeff}.
After data quality requirements, the training data contained $6.5$, $1.1$, and $0.78~\mathrm{mHz}$ of two-, three-, and four-peak events, respectively. The corresponding time separation distributions were inconsistent with the expectation for events happening in the detector volume, implying the presence of backgrounds. In preparation for fitting the data, the same MC simulations used for calibrations were adapted for WIMP recoils of different masses and three background sources, identifiable by their time separation distributions: contaminants on the internal surface of the detector, particle interactions in the gas volume, and accidental coincidences.

\begin{figure}[h]
\centering
  \subfigure {\includegraphics[width=0.95\linewidth]{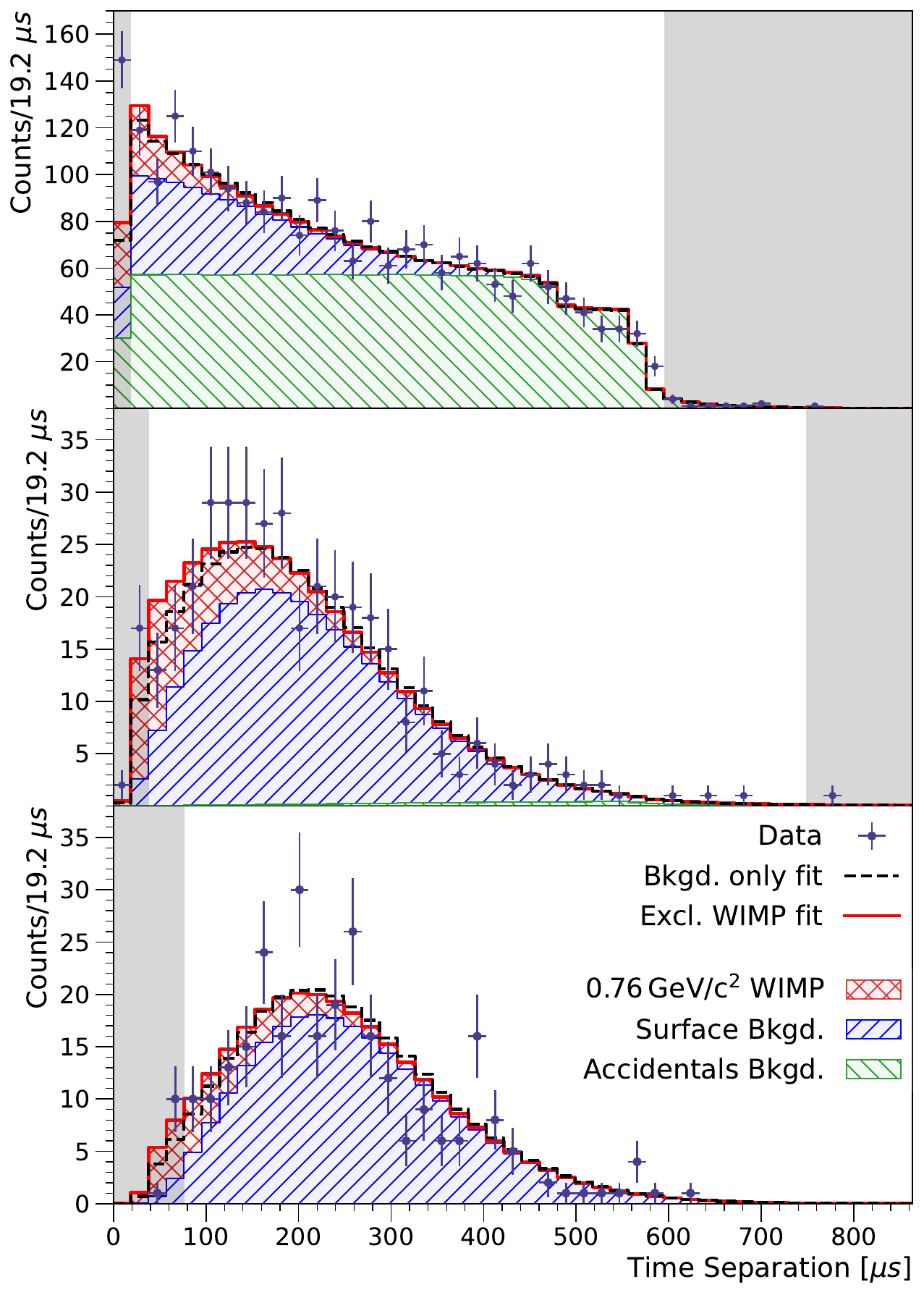}}
  \caption{Top to bottom: Fit (red line) of the time separation distribution of, respectively, two-, three-, and four-peak DM search data (dark blue points), with the stacked contributions from surface (blue hashed) and accidental coincidence (green hashed) backgrounds, together with a $0.76~\mathrm{GeV/c^2}$ WIMP contribution (red hashed) fixed to the 90\% CL\ excluded cross section, $\sigma_{SD_p} = 30.9~\mathrm{pb}$. The best background-only fit (black dashed line), including background volume events, is shown for comparison. The shaded gray areas cover the regions where simulations did not match calibrations, and so were not included in the fit.}
  \label{fig:DataFit}
\end{figure}

For the WIMP elastic scattering recoil energy spectrum, the standard halo parametrization with $\rho_0 = 0.3~\mathrm{GeV/c^2/cm^3}$, $v_0 = 238~\mathrm{km/s}$, $v_\mathrm{Earth} = 232~\mathrm{km/s}$, and $v_\mathrm{esc} = 544~\mathrm{km/s}$ \cite{rave,direct,schnee,phystat} was used, together with the Helm form factor for the nuclear cross section \cite{lewin}, which is nearly identical to $1$. The binding energy of hydrogen to CH$^{+}_{3}$ \cite{NSRDS} was subtracted from the recoil kinetic energy. The rate of WIMP events in the three categories was scaled by the WIMP-proton spin-dependent cross section, $\sigma_{\mathrm{SD}_p}$, which is the parameter of interest.

For surface and volume backgrounds, an underlying energy distribution is necessary to determine the expected relative rates of two-, three-, and four-peak data, and their time separation distributions. It was chosen of the form $R(E) = A + Be^{-E/C}$, where $A$, $B$, $C$ are free parameters for surface and volume contributions separately. The $A$ term represents backgrounds with a uniform energy distribution, such as from Compton interactions; the $Be^{-E/C}$ term is a generic parametrization for rising background rates at low energies, as observed in multiple DM direct detection experiments \cite{10.21468/SciPostPhysProc.9.001} including the previous NEWS-G detector \cite{newsg_first}. Since WIMP recoil also happens in the volume but their energy distribution is fixed by their mass, they form a subclass of volume events, and hence the two contributions are potentially degenerate. To avoid biasing our results, generic volume backgrounds were only considered when generating pseudoexperiments, so that they could accurately represent our observed data; they were instead set to zero when fitting potential WIMP contributions, effectively equivalent to assuming all observed volume events are WIMP recoils. For surface events, the large diffusion time made overlapping electrons less frequent. This, in turn, made the time separation distribution of surface events insensitive to the underlying energy distribution, modifying only the relative rates of observed two-, three-, and four-peak events. Conversely, for volume events, their shorter diffusion time increased the frequency of overlaps and hence the proportion of events with higher number of electrons reconstructed as only having two to four peaks. 

The last background considered was accidental coincidences of unrelated events within $523~\mathrm{\mu s}$ of the trigger time or false positives of the PF method due to baseline noise fluctuations. These were modeled with an MC simulation assuming a uniform time distribution of peaks within the search window, and corroborated by comparing with the distribution of two-, three-, or four-peak events after an $\alpha$ particle, when the increased single-electron event rate produced high rates of coincidences. The rates for each of two-, three-, and four-peak coincident events were left as free independent parameters. The two-, three-, and four-peak data were jointly fit with these four components using the profile likelihood ratio test statistic~\cite{asymptotic}. The fit results on the DM search data are shown in Table~\ref{tab:BackgroundRates}, with an overall background rate of a few $\mathrm{mHz}$. The main contributions were from surface contamination and accidental coincidences, with a smaller contribution from volume background events. 

\begin{table}
\begin{center}
    \begin{tabular*}{\linewidth}{@{\extracolsep{\fill}} l | c  c  c }
\hline \hline
& Two peak & Three peak & Four peak  \\ \hline
Accidental coincidences   & $4.42\pm0.30$ & $0.02\pm0.03$ & $0.00\pm0.01$ \\
Surface background  & $1.43\pm0.29$ & $0.83\pm0.08$ & $0.73\pm0.07$ \\
Volume background   & $0.21\pm0.12$ & $0.15\pm0.08$ & $0.09\pm0.05$ \\
$0.76~\mathrm{GeV/c^2}$ WIMP     & $0.28$ & $0.19$ & $0.11$ \\ \hline \hline
    \end{tabular*}
\protect\caption{Background rates in $\mathrm{mHz}$ obtained from background-only fit on the DM search data. Excluded contribution from $0.76~\mathrm{GeV/c^2}$ WIMP shown for comparison.}
\label{tab:BackgroundRates}
\end{center}
\end{table}


\begin{figure}[t]
\centering
  \includegraphics[width=0.95\linewidth]{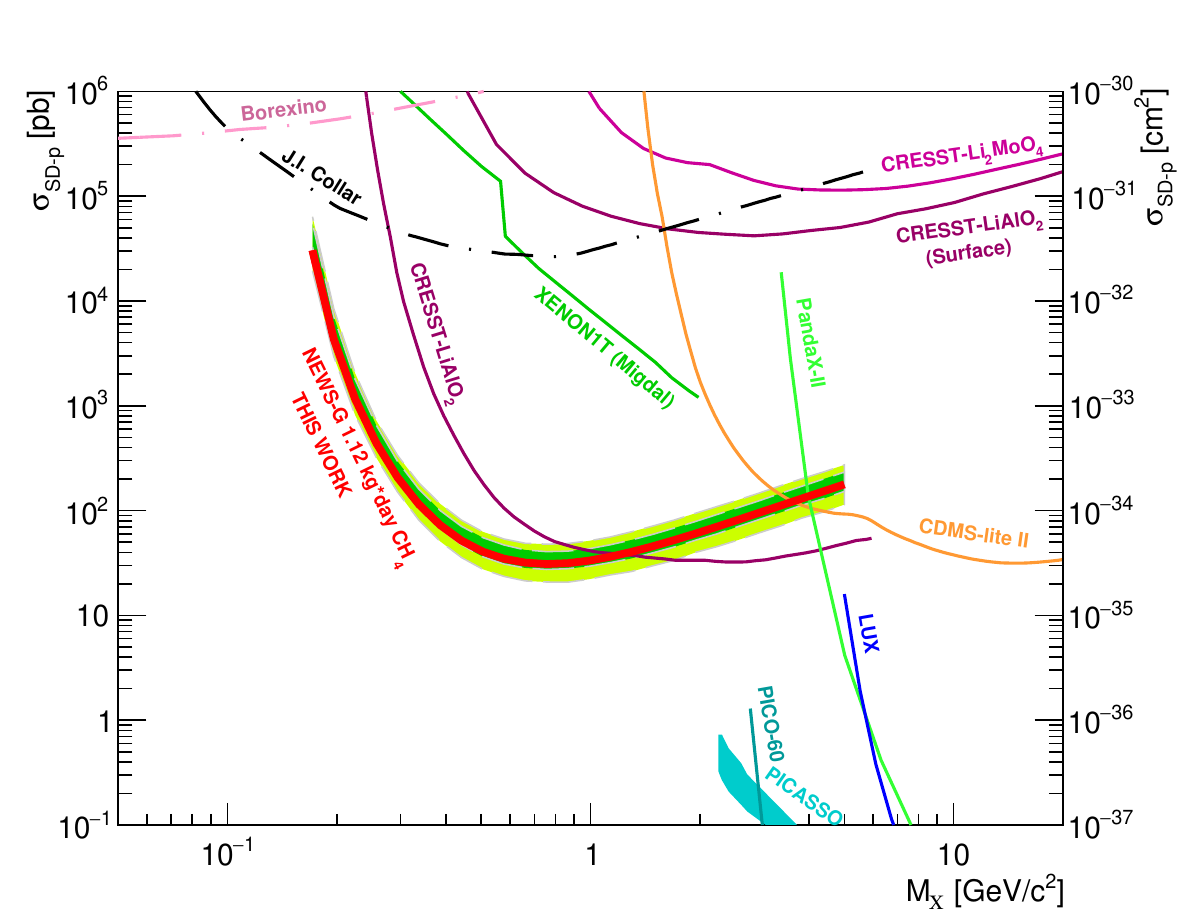}
  \caption{Exclusion limit on the WIMP-proton spin-dependent cross section from this work (thick red line) and 1$\sigma$ and 2$\sigma$ sensitivity bands (dark and light green shaded areas respectively). Upper limits from CDMS-lite \cite{Agnese2018}, CRESST-III \cite{CRESST:2019mle,CRESST:2022jig,CRESST:2022dtl}, LUX \cite{Akerib2017}, PANDAX-II \cite{Xia2019}, XENON-1T (Migdal) \cite{Aprile2019}, PICASSO \cite{Behnke2017}, PICO-60 \cite{Amole2019}, J.I.\ Collar \cite{Collar2018} and Borexino \cite{Bringmann2019} are also shown.}
  \label{fig:ExclusionLimit}
\end{figure}

A constraint on the WIMP-proton spin-dependent cross section was obtained by profiling the likelihood ratio over the rates of the surface and accidental coincidence events; volume background events were fixed to zero due to their near degeneracy with WIMP events. Pseudodatasets were simulated based on the best fit of the training data for various WIMP masses, with cross sections fixed to different values and scaled to the exposure of the DM search data. The distribution of the profile likelihood ratio from these datasets was used to obtain the threshold value of the test statistic for the 90\% CL cross-section exclusion limit. In Fig.~\ref{fig:DataFit}, the fit result is shown for a WIMP with mass $0.76~\mathrm{GeV/c^2}$ and a $\sigma_{\mathrm{SD}_p}$ fixed at $30.9~\mathrm{pb}$ (excluded at 90\% CL). 


The resulting upper limit curve is shown in Fig.~\ref{fig:ExclusionLimit}. New constraints on spin-dependent DM interactions with protons are presented in the mass range $0.17$ to $1.2~\mathrm{GeV/c^2}$, with a  90\% confidence level cross-section upper limit of $30.9~\mathrm{pb}$ for a mass of $0.76~\mathrm{GeV/c^2}$.
Compared to the preliminary training data result \cite{IDM_newsg}, the degree of improvement evolved proportionally to $ 1/\sqrt{\Delta \mathrm{exposure}}$.
Systematic uncertainties from the exposure or selection efficiencies were accounted for by taking their conservative values at $95\%$ CL, and were found to have a negligible effect. The effect of systematic uncertainties on the DM candidate-induced nuclear recoil ionization process was considered through variations of the $W_0$ and $U$ parameters. Negligible effects were observed. The curve in Fig.~\ref{fig:ExclusionLimit} shows the worst-case scenario among all considered combinations. In the context of evaluating the importance of further measurements of the QF, the above result was rederived for the case where the energy threshold is set at the lowest energy for which a QF measurement is available, $510~\mathrm{eV_{nr}}$. In this case, the obtained WIMP sensitivity deteriorates by a factor of 1.4 (13) for a DM mass of $760~\mathrm{MeV/c^2}$ ($300~\mathrm{MeV/c^2}$). 

 
\smallskip

\begin{acknowledgments}
\textit{Acknowledgments}---The help of the technical staff of the Laboratoire Souterrain de Modane is gratefully acknowledged. This research was undertaken, in part, thanks to funding from the Canada Excellence Research Chairs Program, the Canada Foundation for Innovation, the Arthur B. McDonald Canadian Astroparticle Physics Research Institute, Canada, the French National Research Agency (ANR-15-CE31-0008), and the Natural Sciences and Engineering Research Council of Canada. This project has received support from the European Union's Horizon 2020 research and innovation program under grant agreements no.~841261 (DarkSphere), no.~845168 (neutronSPHERE), and no.~101026519 (GaGARin).
Support from the U.K. Research and
Innovation — Science and Technology Facilities Council (UKRI-STFC), through grants no.~ST/V006339/1, no.~ST/S000860/1, no. ST/W000652/1, no. ST/X005976/1, and no. ST/X508913/1, the UKRI Horizon Europe Underwriting scheme (GA101066657/Je-S EP/X022773/1), and the Royal Society International Exchanges Scheme (IES$\backslash$R3$\backslash$170121) is acknowledged.
Support by the Deutsche Forschungsgemeinschaft (DFG, German Research Foundation) under Germany’s Excellence Strategy — EXC 2121 “Quantum Universe” — 390833306 is acknowledged.
\end{acknowledgments}

\bibliography{bib_lsm_letter}

\onecolumngrid
\appendix
\section{End Matter}
\twocolumngrid

\setcounter{equation}{0}
\renewcommand{\theequation}{A\arabic{equation}}

\textit{Appendix A: Polya and COM-Poisson distributions}---The stochastic fluctuations in the production of ionization electrons in the gas is characterized by the Fano factor $F$ \cite{uno2}, defined as the variance divided by the mean number of electrons. For a Poisson-distributed process $F=1$, but for many ionization detectors it is known that $F$ is $ \lesssim 0.2$ \cite{scinprop,policarpo,owens,germanium}, indicating less dispersion in the number of ionization electrons. There is no physically motivated, \emph{a priori} probability distribution function (PDF) to model this type of dispersion. Therefore, the NEWS-G collaboration utilizes the COM-Poisson (COnway Maxwell Poisson) distribution \cite{queuing, mainCOM} for this purpose. It is a suitable, known distribution that can be adapted to give a discrete PDF with any value of $F$ for any mean number of ionization electrons $\mu$ \cite{newsg_fano}, allowing for proper treatment of $F$ as a systematic. The PDF is defined in terms two parameters, $\lambda$ and $\nu$ \cite{usefulCOM},

\begin{equation}
    P(X = x|\lambda,\nu) = \frac{\lambda^x}{(x!)^{\nu} Z(\lambda,\nu)},
\end{equation}

\noindent for $x\, \in\, \mathbb{N}_0, \;\lambda > 0,\; \nu \geq 0$. $Z$ is a normalization factor,

\begin{equation}
    Z(\lambda,\nu) = \sum_{s=0}^{\infty} \frac{\lambda^s}{(s!)^{\nu}}.
\end{equation}

While the PDF parameters $\lambda$ and $\nu$ predominantly control the mean and dispersion of the PDF, respectively, these parameters do not correspond to $\mu$ and $F$, and no closed-form expression relating them exists. Numerical lookup tables to overcome this obstacle have been developed and are publicly available \cite{newsg_fano, newsg_website}. This model has been successfully used to characterize $\mathrm{^{37}Ar}$ calibration data for NEWS-G in various gas conditions, including pure $\mathrm{CH_4}$ \cite{Arora:2024tbs, newsg_laser}.

The second process that defines the energy response of spherical proportional counters---avalanche amplification---has been historically described by the Polya PDF  \cite{L11,L12,L13,L14,newsg_laser}. The PDF for the avalanche size $A$ from a single gas ionization electron is given as \cite{newsg_laser}

\begin{equation}
\begin{aligned}
    P_{\mathrm{Polya}}\left(A | \theta, \left \langle G \right \rangle \right) & = \frac{1}{\left \langle G \right \rangle} \frac{\left(1 + \theta \right)^{1 + \theta}}{\Gamma \left(1 + \theta \right)} \left( \frac{A}{\left \langle G \right \rangle} \right)^{\theta} \\ & \times \exp \left( -\left(1 + \theta \right)\frac{A}{\left \langle G \right \rangle} \right),
\end{aligned}
\label{polya_eq}
\end{equation}

\noindent characterized in this context by a shape parameter $\theta$ and the mean gain of the detector $\left \langle G \right \rangle$. The latter quantity is defined as the average number of avalanche electrons produced from a single gas ionization electron \cite{newsg_laser}. The $\theta$ parameter can modify the shape of this distribution from exponential ($\theta = 0$) to Gaussian ($\theta \gg 1$); for NEWS-G proportional counters, $\theta$ is typically $< 1$ \cite{newsg_laser, Arora:2024tbs}.

\textit{Appendix B: Detector stability}---In addition to daily low-intensity laser calibration data, high-intensity laser data was taken continuously throughout the physics campaign. The laser was operated at a pump diode current of $140~\mathrm{A}$ instead of $113$--$120~\mathrm{A}$, resulting in a factor of $\mathcal{O}(100)$ increase in the number of extracted photoelectrons. Because of the large amplitude of these events in the photodiode channel, they could be tagged with presumed $100\%$ efficiency. The utility of this calibration data was to provide real-time monitoring of changes in the detector response over time. 

\begin{figure}
  \centering
  \includegraphics[width=0.48\textwidth]{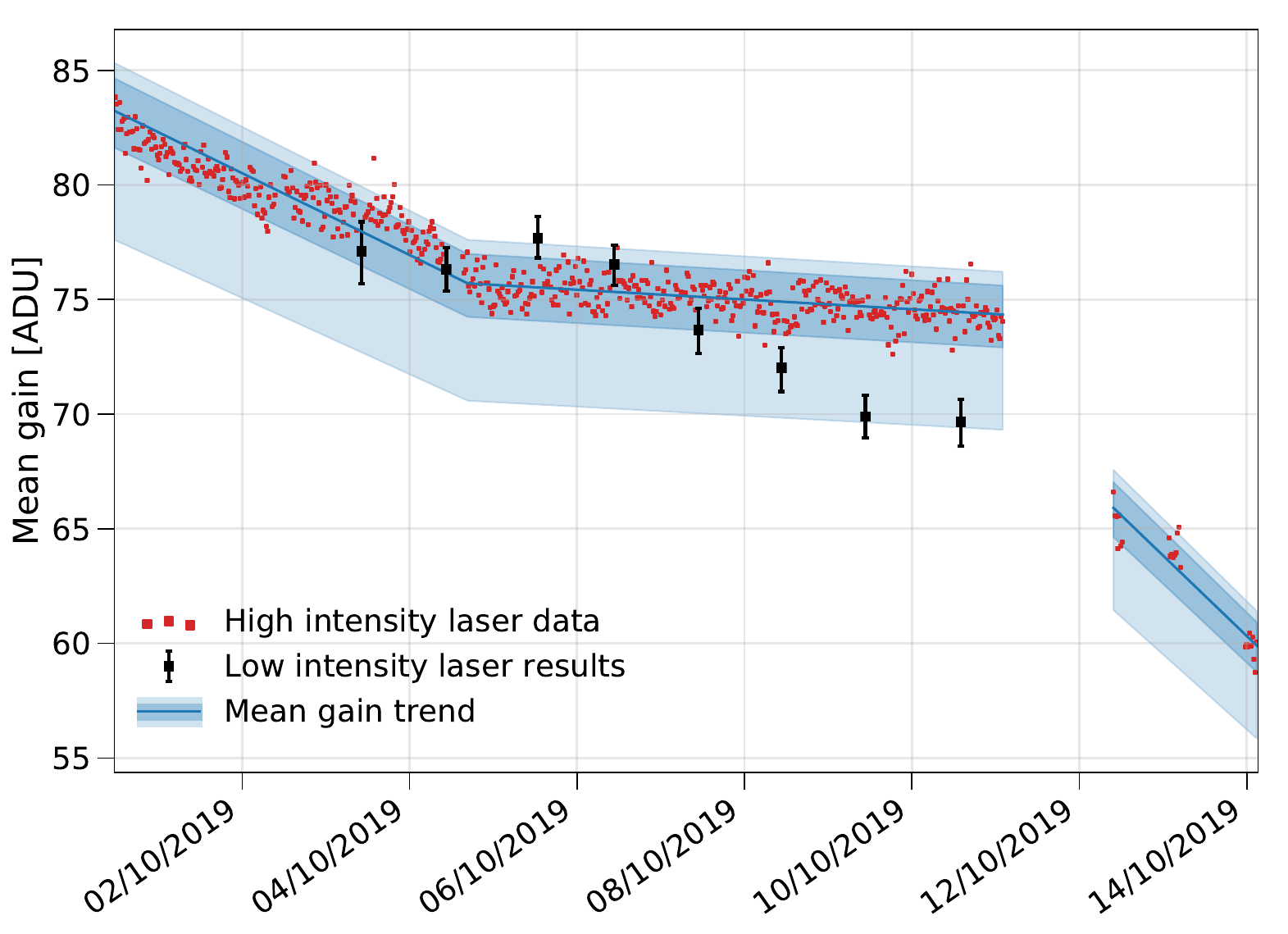}
  \caption{\label{fig:GainTrend} The trend of high-amplitude laser calibration data (the ratio of sphere to photodiode pulse amplitude) over time is shown as red markers. This trend was fit with a piecewise linear function, shown with $1 \sigma$ and $2 \sigma$ error bars (blue line and shaded regions). This trend was then scaled by the absolute gain measurements of all low-intensity laser calibration runs (orange markers with $1 \sigma$ statistical error bars). The gap of time around October $10$ not fit by the piecewise linear function was omitted due to the presence of a neutron calibration source, which significantly changed the charge extraction efficiency of the UV laser.}
\end{figure}

Increasing charge attachment was one such effect that was expected, as continuous outgassing of electronegative contaminants can occur inside the SPC \cite{newsg_first}. This results in a decrease in the apparent amplitude of the laser calibration events observed (of a fixed intensity). This can be accounted for in the overall detector response model as a time dependence of the mean gain of the detector discussed in the previous subsection, $\left \langle G \right \rangle$. However, a confounding effect that would yield a similar observation is the aging of the fiber-optic cable directing the UV laser light into the SPC. This is a known problem, especially in the case of UV lasers \cite{newsg_laser}. To disentangle these two effects, the ratio of the sphere and photodiode amplitude over time is used, dividing out the fiber aging, as it affects identically both the sphere and photodiode signals. The resulting trend is shown in Fig.\ \ref{fig:GainTrend}.

As expected, there is a general decline of the UV laser signal over time due to increasing charge attachment. The exact reason for the change in the rate of decrease around October $5$ is not known. The more significant change after October $12$ was due to the introduction of a canister containing $\mathrm{^{37}Ar}$, bringing with it gas contaminants.
This trend was fit with a three-part piecewise linear function (with floating knot points), omitting a period of time where a neutron calibration source introduced into the SPC significantly changed the charge extraction efficiency of the detector. Next, to adopt this trend as a time dependence for $\left \langle G \right \rangle$, the ratio over time was multiplied by a single scaling factor, which was fit to the low-intensity laser calibration results for $\left \langle G \right \rangle$. The resulting scaled trend is shown in Fig.\ \ref{fig:GainTrend}, showing an $11\%$ drop in $\left \langle G \right \rangle$ between the beginning and end of the physics campaign ($27\%$ by the end of the $\mathrm{^{37}Ar}$ calibrations). For the subsequent DM analysis, the conservative, simplifying choice was made to adopt the value of $\left \langle G \right \rangle$ from this trend at the time of the last physics dataset on October $10$, when the gain was the lowest.

\begin{figure}
\centering 
\includegraphics[width=0.45\textwidth]{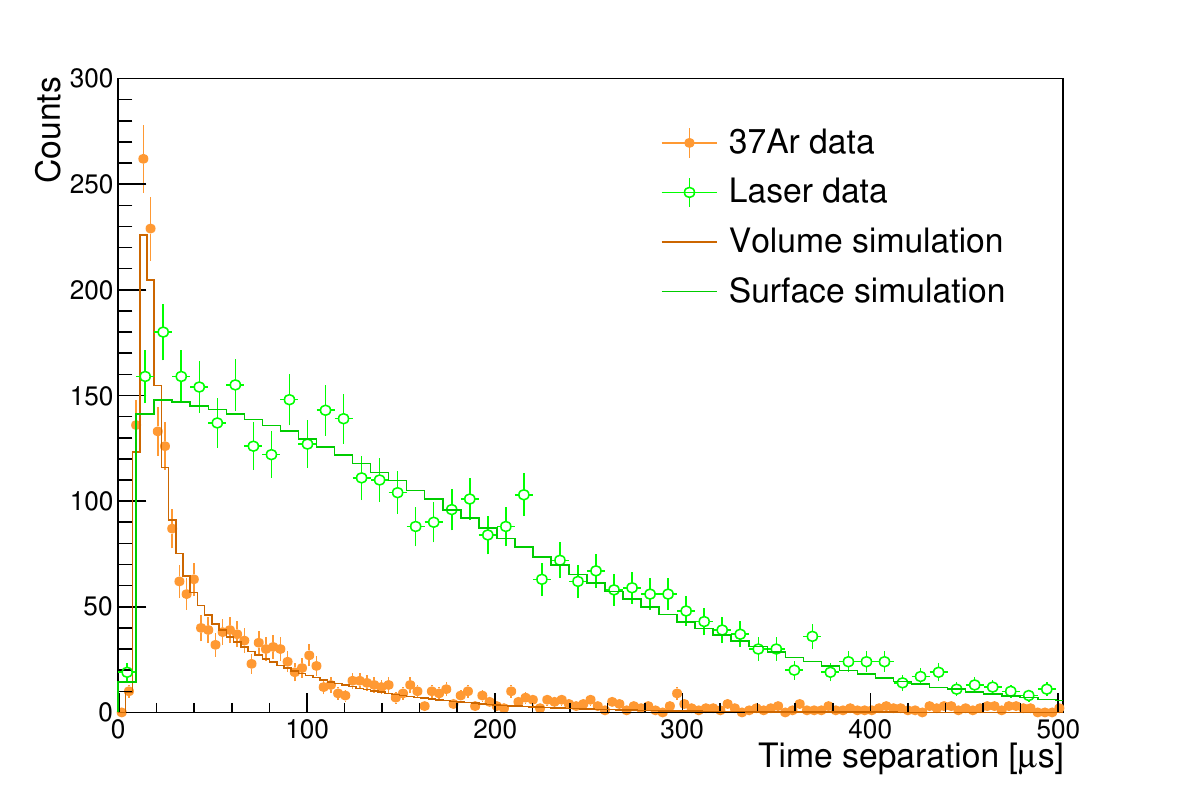}
\caption{Comparison between the time separation spectra of calibrations and simulations for two-peak data.}
\label{fig:PF_calib_comp}
\end{figure}

\textit{Appendix C: Peak and time separation detector response model}---A model is required to determine the probability that an event with $n_{e^-}$ ionization electrons will be reconstructed by the PF algorithm as having $k_{p}$ peaks. This will also depend on the mean diffusion time they experience, $\sigma_{\mathrm{diff}}$, since electrons collected close enough in time cannot be distinguished by the PF algorithm . Furthermore, for a given $n_{e^-}$ and $\sigma_{\mathrm{diff}}$, the PDF for the reconstructed peak time separation is also required, to be able to discriminate against surface backgrounds. An \emph{ad hoc} model was developed, based on results from laser and $^{37}$Ar calibrations; for an event with $n_{e^-}$ ionization electrons, the model works as follows.

First, $n_{e^-}$ arrival times are generated. If the simulated event is tagged as ``volume,'' a random radius is drawn following a $\rho(r) \propto r^2$ distribution (due to the detector's spherical geometry), which is converted into a mean diffusion time based on the diffusion model $\sigma_{\mathrm{diff}}(r) = \sigma_{\mathrm{max}}\times (r/r_{\mathrm{max}})^\alpha$, where $\sigma_{\mathrm{max}} $ and $ \alpha$ are derived from calibrations; for surface events,  $\sigma_{\mathrm{max}}$ is used directly. Some electrons are removed randomly to represent attachment, with a probability each of $\mathrm{att}(r) = \mathrm{att}_{\mathrm{max}} \times \sigma_{\mathrm{diff}}(r) / \sigma_{\mathrm{max}}$, where $\mathrm{att}_{\mathrm{max}}$ is based on $^{37}$Ar calibrations. The probability for the event to trigger and be tagged as far-side only is also applied here, based on the total number of electrons after attachment and their starting radius.

Second, proceeding left to right (forward in time) in the list of arrival times, the electrons are split into ``bundles'' of overlapping peaks, by using the error function obtained from the laser calibration to determine the probability that the PF method can separate them. Each bundle is associated with a certain number of found peaks based on its size. For single-electron bundles, the PF's efficiency for isolated electrons determines the probability of finding one or zero peaks. For bundles with exactly two electrons, it is assumed that the PF method always finds the feature and identifies it as one single peak. For bundles with more than two electrons, where consecutive peaks cannot be split, there is a chance the PF method can separate some of the nonconsecutive ones (e.g., first and second peaks are too close in time to be split, and so are the second and third, but first and third peaks are not). To determine whether nonconsecutive peaks can be split, the same error function used for two nonoverlapping peaks is used, but with an additional \emph{ad hoc} penalty term of $50\%$ (obtained by comparison with $^{37}Ar$ calibration data). The number of peaks found in the bundle is one plus the number of nonconsecutive peaks split this way.

This produces the probability to observe an  $n_{e^-}$-electron event as having $k_{p}$ observed peaks with a time separation of $\Delta t_{\mathrm{peaks}}$ between the first and last found peak or bundle. Arrival times are repeatedly drawn and passed to this procedure until reaching the desired statistical uncertainty on the final distribution. The results from this model are compared with calibrations in Fig.~\ref{fig:PF_calib_comp}. Since only $k_{p}$ is available in true data, not $n_{e^-}$, this model's predictions depend on the source's underlying distribution of $n_{e^-}$ (or, equivalently, ionization energy).

\end{document}